\documentclass[a4paper,11pt]{article}
\usepackage{pos}
\usepackage{graphicx}
\usepackage{amsmath}
\usepackage{braket}

\title{Next-to-leading power gluon TMDs from back-to-back DIS dijets at next-to-eikonal accuracy at low x}
\author{Tolga Altinoluk}
\author{Guillaume Beuf}
\author{Alina Czajka}
\author*{Kacper Gosławski}
\affiliation{National Centre for Nuclear Research,\\
  Pasteura 7, Warsaw 02-093, Poland}

\emailAdd{tolga.altinoluk@ncbj.gov.pl}
\emailAdd{guillaume.beuf@ncbj.gov.pl}
\emailAdd{alina.czajka@ncbj.gov.pl}
\emailAdd{kacper.goslawski@ncbj.gov.pl}

\abstract{We calculate next-to-leading power contributions to the gluon TMDs in DIS dijet production in the back-to-back limit at low x within the Color Glass Condensate (CGC) effective field theory at next-to-eikonal accuracy. We put a special emphasis on three-point correlation functions from CGC calculations, including their correspondence to an appropriately chosen definition of three-point TMDs within the aforementioned kinematical conditions. We also discuss importance of time ordering as well as momentum fraction space formulation of the correlators. 
Finally, we calculate the dijet cross section corresponding to the three-point CGC correlator as a combination of TMD functions.}

\FullConference{
}

\begin{document}
\maketitle

\section{Introduction}
The Color Glass Condensate (CGC) effective field theory provides a well-established theoretical framework suitable for description of highly boosted dilute-dense systems, such as Deep Inelastic Scattering, in the Regge-Gribov or small-x limit \cite{CGC}. In this limit, the effect of gluon saturation renders a possibility of using the semi-classical approximation, in which the dense gluonic target may be viewed as an ensemble of classical background color fields while the projectile is treated perturbatively.

One of the most frequently used approximations in the CGC formalism, known as the eikonal approximation, is equivalent to taking into account only terms with leading power of energy. At the level of the background field it corresponds to 
\begin{equation}
A^\mu(x^-, x^+, \mathbf{x}_\perp) \approx \delta^{\mu-}\delta(x^+)A^-(\mathbf{x}_\perp),
\end{equation}
which arises due to a specific hierarchy of background field components in a boosted reference frame,
\begin{equation}
A^- = \mathcal{O}(\gamma) \gg A_j  = \mathcal{O}(1) \gg A^+ = \mathcal{O}(1/\gamma),
\end{equation}
and due to Lorentz contraction and time dilation. In next-to-eikonal (Neik) corrections, adopted in all of our calculations,  all of these constraints have been relaxed. 

In essence, the interaction of the projectile (quark or anti-quark in the presented case) with the dense gluonic target translates to the scattering off a set of background color field insertions. The Next-to-eikonal corrections to the background quark propagator has been already established in the literature \cite{beuf-1}.

\begin{figure}[htbp]
    \centering
    \includegraphics[width=0.75\textwidth]{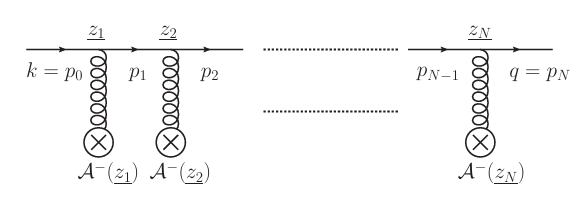}
    \caption{Quark scattering on $A^-$ background gluon field insertions, as can be seen in \cite{beuf-1}.}
    \label{fig:propagator}
\end{figure}

In order to bridge the CGC calculations with the standard TMD formalism, we consider dijet production in Deep Inelastic Scattering via exchanged of a virtual photon
\begin{equation}
    \label{dijet-general-cross-section}
    \frac{d\sigma^{l+\mathrm{target}\to l^{\prime}+\mathrm{dijet}+X}}{dx_{Bj}\, dQ^{2}\, dP.S.}
    =
    \frac{\alpha_{em}}{\pi\, x_{Bj}\, Q^{2}}
    \left[
        \left( 1 - y + \frac{y^{2}}{2} \right)
        \frac{d\sigma_{\gamma^{*}_{T}\to \mathrm{dijet}}}{dP.S.}\left(x_{Bj}, Q^{2}\right)
        +
        \left( 1 - y \right)
        \frac{d\sigma_{\gamma^{*}_{L}\to \mathrm{dijet}}}{dP.S.}\left(x_{Bj}, Q^{2}\right)
    \right],
\end{equation}
where $\gamma_T$ and $\gamma_L$ stand for transversely and longitudinally polarized photon, $x_{B_j}$ is the Bjorken x, and $y$ is the inelasticity variable. The cross-section is then calculated within the dipole approximation, where the process can be regarded as splitting of the incoming photon into a quark-antiquark pair outside or inside the target. This goal also has been achieved, using the aforementioned next-to-eikonal background propagator \cite{beuf-2}.
\begin{figure}[htbp]
    \centering
    \includegraphics[width=1\textwidth]{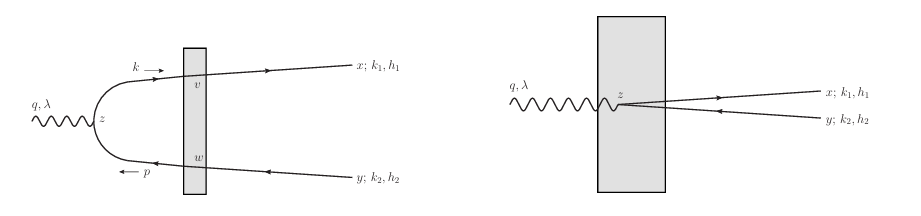}
    \caption{Next-to-eikonal contributions to the dijet production cross section with photon splitting outside (left) and inside (right) the target. Figure taken from \cite{beuf-2}.}
    \label{fig:crosssection}
\end{figure}

The high energy limit of the TMD distributions might be obtained from next-to-eikonal CGC calculations within so called correlation limit, where both jets are created nearly back-to-back. Defining the total transverse momentum of the jets as $|\mathbf{k|}$ and the relative transverse momentum between the jets as $|\mathbf{P|}$, the correlation regime implies that $|\mathbf{k| \ll |P|}$. Perturbatively, the power corrections are of the form
\begin{equation}
     \bigg(\frac{|\mathbf{k|}}{|\mathbf{P}|}\bigg)^n,
\end{equation}
with $n = 0$ corresponding to the leading power contributions, $n = 1$ to the next-to-leading power (NLP) contributions and so forth \cite{beuf-3}. Fourier transformation of these momenta variables leads to an equivalent statement of the correlation regime while $|\mathbf{r}| \ll \mathbf{|b|}$, where $\mathbf{r}$ is the transverse dipole size and $\mathbf{b}$ is the impact parameter.

Our goal in the current work is to find a mapping between TMD distributions in the back-to-back limit and CGC correlators of the classical field insertions at next-to-eikonal level using the aforementioned cross-section, calculated in \cite{beuf-2}, perturbatively expanded up to NLP corrections.

\section{General procedure}
The cross section for the longitudinal photon case may be separated into strictly eikonal part and an additional next-to-eikonal correction \cite{beuf-3}:
\begin{equation}
    \frac{d\sigma_{\gamma^* \rightarrow q_1\bar{q}_2}}{dP.S.} = \frac{d\sigma_{\gamma^*\rightarrow q_1\bar{q}_2}}{dP.S.} \Bigg|_{Eik} + \frac{d\sigma_{\gamma^*\rightarrow q_1\bar{q}_2}}{dP.S.}\Bigg|_{NEik}.
\end{equation}
Here the eikonal part reads:
\begin{equation}
    \frac{d\sigma_{\gamma^*\rightarrow q_1\bar{q}_2}}{dP.S.} \Bigg|_{Eik} = (2q^+)2\pi \delta(k^+_1 + k^+_2 - q^+)\sum_{hel., col.}\Big<\big(\mathbf{M}^{Eik}_{\gamma^*_L \xrightarrow{}q_1\bar{q_2}}\big)^\dagger\mathbf{M}^{Eik}_{\gamma^*_L \xrightarrow{}q_1\bar{q_2}} \Big>,
\end{equation}
while the next-to-eikonal correction is
\begin{equation}
    \frac{d\sigma_{\gamma^*\rightarrow q_1\bar{q}_2}}{dP.S.} \Bigg|_{NEik} = (2q^+)2\pi \delta(k^+_1 + k^+_2 - q^+)\sum_{hel., col.}2Re\Big<\big(\mathbf{M}^{Eik}_{\gamma^*_L \xrightarrow{}q_1\bar{q_2}}\big)^\dagger\mathbf{M}^{NEik}_{\gamma^*_L \xrightarrow{}q_1\bar{q_2}} \Big>.
\end{equation}
All the constituents of the eikonal and next-to-eikonal amplitudes $\mathbf{M}_{\gamma^*_L \xrightarrow{}q_1\bar{q}_2}$ have a similar operational structure:
\begin{equation}
    \mathcal{O} = \int_{\mathbf{r}, \mathbf{b}}e^{-i\mathbf{r}\cdot\mathbf{P}}e^{-i\mathbf{b}\cdot\mathbf{k}}f(\mathbf{r})U(\mathbf{b}, \mathbf{r}, b^-),
\end{equation}
where $f(\mathbf{r})$ is a certain numerical function, and $U(\mathbf{b}, \mathbf{r}, b^-)$ is an overall combination of Wilson lines depending on the operator under consideration. For example, the eikonal operator reads:
\begin{equation}
    \mathcal{O}^{eik} = \int_{\mathbf{r}, \mathbf{b}}e^{-i\mathbf{r}\cdot\mathbf{P}}e^{-i\mathbf{b}\cdot\mathbf{k}}f(\mathbf{r})\big[U_F(\mathbf{b} + z_2\mathbf{r}, b^-)U_F(\mathbf{b} - z_1\mathbf{r}, b^-)^\dagger - 1\big].
\end{equation}
The next-to-eikonal operators are expressed in the same fashion, albeit they are much more algebraically involved. As all of them are treated alike, the structurally simpler eikonal operator has been set as an example of the procedure stated below.

The Wilson line structure is expanded in powers of $|\mathbf{r|}$ up to the desired order. For the aforementioned eikonal operator the expansion is \cite{beuf-3}:
\begin{equation}
\begin{split}
U_F(\mathbf{b}+z_2\mathbf{r},b^-)\,&U_F(\mathbf{b}+z_1\mathbf{r},b^-)^{\dagger}-1
    = \left(1+i\frac{z_2-z_1}{2}\,\mathbf{r}\cdot\mathbf{k}\right)\,\mathbf{r}^{i}\,\partial_{i}U_F(\mathbf{b},b^-)\,U_F(\mathbf{b},b^-)^{\dagger}\\
    &-\frac{1}{2}\,\mathbf{r}^{i}\mathbf{r}^{j}\left[\partial_{i}U_F(\mathbf{b},b^-)\,\partial_{j}U_F(\mathbf{b},b^-)^{\dagger}\right]
    +\mathcal{O}\!\left(\frac{|\mathbf{r}|^{3}}{|\mathbf{b}|^{3}}\right)
    \label{eh},
\end{split}
\end{equation}
with the Wilson line derivative $\partial_jU_F(\mathbf{b}, b^-)$ defined as
\begin{equation}
    \partial_j U_F(\mathbf{b}, b^-) = - \int_{-\infty}^{\infty} \mathrm{d}z^{+}\, U_F(\infty, z^{+}; \mathbf{b}, b^-) \left( i g F^{-}_{j}(z) \right) U_F(z^{+}, -\infty; \mathbf{b}, b^-)
    \label{eq:placeholder_label}
\end{equation}
and $z_1$, $z_2$ being the light cone + momentum fractions carried by the jets \cite{beuf-3}. Contributions with one or two field strength insertions are included in eq. \eqref{eh} and in all next-to-eikonal operators as well, at amplitude level. Inserting these expanded formulas and squaring the amplitude, we can regroup the dijet cross section as
\begin{equation}
    \frac{d\sigma_{\gamma_L^* \rightarrow q_1\bar{q}_2}}{dP.S.}\Bigg|_{Eik +Neik} = \frac{d\sigma_{\gamma_L^* \rightarrow q_1\bar{q}_2}}{dP.S.}\Bigg|_{F^{{\perp}-}F^{{\perp}-}} + \frac{d\sigma_{\gamma_L^* \rightarrow q_1\bar{q}_2}}{dP.S.}\Bigg|_{F^{{\perp}-}F^{{\perp}-}F^{{\perp}-}} + \frac{d\sigma_{\gamma_L^* \rightarrow q_1\bar{q}_2}}{dP.S.}\Bigg|_{F^{+-}F^{{\perp}-}},
\end{equation}
where every part consists of different number and types of the $F$ field insertions. For example, the two-point gluon field insertion term is, up to a constant factor and keeping the Wilson lines implicit,
\begin{equation}
    \frac{d\sigma_{\gamma_L^* \rightarrow q_1\bar{q}_2}}{dP.S.}\Bigg|_{F^{{\perp}-}F^{{\perp}-}} \sim g^2\int_{\mathbf{b}_1,\mathbf{b}_2}e^{i(\mathbf{b}_1 - \mathbf{b}_2) \cdot \mathbf{k}}\int dz_1^+dz_2^+ \big<F_j^-(z_2^+, \mathbf{b}_2)F_i^-(z_1^+, \mathbf{b}_1)\big> \cdot \Pi^{ij}(\mathbf{P}, \mathbf{k}, \{z\}).
\end{equation}
Here the $\Pi^{ij}(\mathbf{P}, \mathbf{k}, \{z\})$ is a function of dijet momenta and the insertions reside at two different transverse points. The properties of the two-point function, together with their connection to the TMD formalism, has been studied in \cite{beuf-3}.

\section{The three-point function}
The three insertions part of the cross section,
\begin{equation}
\begin{split}
    \frac{d\sigma_{\gamma_L^* \rightarrow q_1\bar{q}_2}}{dP.S.}\Bigg|_{3*F^{{\perp}-}} \sim g^3&\int_{\mathbf{b}_1,\mathbf{b}_2}e^{i(\mathbf{b}_1 - \mathbf{b}_2) \cdot \mathbf{k}}\int dz_1^+dz_2^+dz^+_3 \big<F_k^-(z_3^+, \mathbf{b}_2)F_j^-(z_2^+, \mathbf{b}_2)F_i^-(z_1^+, \mathbf{b}_1)\big> \\& \cdot \Pi^{ijk}(\mathbf{P}, \{z\}),
\end{split}
\end{equation}
still depends only on two different transverse points. Its connection to the TMD distributions, similarly to the two-point case, may be established by the identification of the classical color average with an appropriate quantum TMD operator \cite{TMDconn}
\begin{equation}
    \big<\mathcal{O}\big> \rightarrow \lim_{P' \rightarrow P} \frac{\bra{P'} \hat{\mathcal{O}} \ket{P}}{\braket{P'|P}}.
\end{equation}
The normalization is taken as
\begin{equation}
    \braket{P'|P} = 2P'(2\pi)^3\delta(P^{'-} - P^-)\delta^{(2)}(\mathbf{P}' - \mathbf{P}).
\end{equation}
As the procedure here is ambiguous due to importance of operator ordering in the formalism, the properly established time order of fields in the case of the three-point function needs to be taken with care, especially that all the calculations up to the amplitude level involve integration over whole phase space. This issue can be avoided with adequate symmetrization in the process. We can also define the following time-ordered gluonic operators:
\begin{equation}
    \mathcal{O}^i_a(z^+, \mathbf{z}) = T\big[U_A(\infty, z^+, \mathbf{z})_{aa'}F^i_{a'}(z^+, \mathbf{z})\big],
\end{equation}
\begin{equation}
    \mathcal{O}^{ij}_{ab}(z_1^+, z^+_2, \mathbf{z}) = T\big[U_A(\infty, z_1^+, \mathbf{z})_{aa'}F^i_{a'}(z_1^+, \mathbf{z})U_A(\infty, z_2^+, \mathbf{z})_{bb'}F^j_{b'}(z_2^+, \mathbf{z})\big].
\end{equation}
The second definition is symmetric under total exchange of all indices and first two arguments - $\mathcal{O}^{ij}_{ab}(z_1^+, z^+_2, \mathbf{z}) = \mathcal{O}^{ji}_{ba}(z_2^+, z^+_1, \mathbf{z})$. This allows us to write the three-point operator in the following form:
\begin{equation}
\begin{split}
    \frac{d\sigma_{\gamma_L^* \rightarrow q_1\bar{q}_2}}{dP.S.}\Bigg|_{3*F^{{\perp}-}} \sim g^3&\int_{\mathbf{b}_1,\mathbf{b}_2}e^{i(\mathbf{b}_1 - \mathbf{b}_2) \cdot \mathbf{k}}\int dz_1^+dz_2^+dz^+_3 \big<\mathcal{O}^c_k(z^+_1, \mathbf{b}_\perp + \Delta\mathbf{\mathbf{b}})\mathcal{O}^{ab}_{ij}(z^+_1, z^+_2, \mathbf{b})\big> \\ &\cdot (d^{abc}[\dots] + f^{abc}[\dots]),
\end{split}
\end{equation}
where we divided the momentum coefficient into symmetric and antisymmetric parts. Experimentally, it is very challenging to distinguish light quark or antiquark (or gluon) jets. At the level of dijet cross section with unidentified flavors, the $d^{abc}$ terms in our result cancel out, since they are antisymmetric by exchange of the produced quark and antiquark. This leaves us with
\begin{equation}
\begin{split}
    \frac{d\sigma_{\gamma_L^* \rightarrow \textrm{dijet}}}{dP.S.}\Bigg|_{3*F^{{\perp}-}} \sim g^3&\int d^2\mathbf{b} e^{i\Delta\mathbf{b} \cdot \mathbf{k}}\int dz_1^+dz_2^+dz^+_3 \big<\mathcal{O}^c_k(z^+_1, \mathbf{b}_\perp + \Delta\mathbf{\mathbf{b}})\mathcal{O}^{ab}_{ij}(z^+_1, z^+_2, \mathbf{b})\big> \\ &\cdot f^{abc}\frac{(z_1-z_2)(b_2^+-b^+_3)}{2q^+z_1z_2} \frac{4\mathbf{P}^i\mathbf{P}^j\mathbf{P}^k}{(\mathbf{P}^2+\bar{Q}^2)^4}.
\end{split}
\end{equation}
To compare our result with a suitable three-point TMD gluon distribution, we utilize the definitions described in \cite{vladimirov}. Adopted to our notation, these are
\begin{equation}
  \Phi^{g;cab}_{12;\,k i j}\!\left(z_1^{+}, z_2^{+}, z_3^{+}, \Delta \mathbf{b}_{\perp}\right)
  = \bra{P} g\,O_{k}^{c}\left(z_1^{+},\Delta \mathbf{b}_{\perp}\right)^{\dagger}
  O_{ij}^{ab}\!\left(z_3^{+}, z_2^{+}, \mathbf{0}_{\perp}\right) \ket{P}.
\end{equation}
With our prescription, the operator part takes the form:
\begin{equation}
    \int dz^+_1 \int d^2\mathbf{b} \big<\mathcal{O}^c_k(z^+_1, \mathbf{b}_\perp + \Delta\mathbf{\mathbf{b}})\mathcal{O}^{ab}_{ij}(z^+_1, z^+_2, \mathbf{b}\big> = \frac{1}{2P^-} \Phi^{g;cab}_{12;\,k i j}\!\left(0^{+}, \Delta z_2^{+}, \Delta z_3^{+}, \Delta \mathbf{b}_{\perp}\right)
\end{equation}
with $\Delta z_n^+ = z_n^+ - z_1^+$. It is possible to transform this operator to the momentum fraction space \cite{vladimirov}:
\begin{equation}
    \int dz^+_1 \big<\mathcal{O}^c_k(z^+_1, \mathbf{b}_\perp + \Delta\mathbf{\mathbf{b}})\mathcal{O}^{ab}_{ij}(z^+_1, z^+_2, \mathbf{b}\big> = \frac{(P^-)^2}{2}\int[dx]e^{-i(x_2^+\Delta z^+_2 + x^+_3\Delta z^+_3) P^-}\Phi^{g;cab}_{12;\,k i j}\!\left(x^+_1, x^+_2, x^+_3, \Delta \mathbf{b}_{\perp}\right),
\end{equation}
where the measure is
\begin{equation}
    \int [dx] = \int_{-1}^1 dx^+_1dx^+_2dx^+_3\delta(x^+_1 + x^+_2 + x^+_3).
\end{equation}
The full three-point function at next-to-leading power, next-to-eikonal accuracy, written down using the above TMD definition, reads (keeping $d^{abc}$ terms for completeness): 
\begin{equation*}
\begin{split}
\frac{d\sigma_{\gamma^*_L \rightarrow q_1\bar{q}_2}}{dP.S.}&\Bigg|_{3*F^{\perp-}} = (2q^+)2\pi\delta(p^+ - k^+)4Q^2e^2e_f^2z^3(1-z)^3\int d^2\mathbf{b} \, e^{i\Delta\mathbf{b}\cdot\mathbf{k}} \frac{(P^-)^2}{2} \\ & \cdot \int [dx]\Phi^{g;cab}_{12;ijk}(x_1,x_2,x_3, \Delta \mathbf{b})\int d\Delta z_2^+ \int d\Delta_z3^+\,e^{-i(x_2\Delta z^+_2 + x_3 \Delta z_3^+)P^-} \\ & \cdot\Bigg\{f^{abc}\frac{(1-2z)\mathbf{P}^i\mathbf{P}^j\mathbf{P}^k}{z(1-z)q^+[\mathbf{P}^2+\bar{Q}^2]^4}[\Delta z^+_2 -\Delta z_3^+] +d^{abc}\bigg[\frac{\delta^{ij}}{[\mathbf{P}^2+\bar{Q}^2]^4}-\frac{4\mathbf{P}^i\mathbf{P}^j\mathbf{P}^k}{[\mathbf{P}^2+\bar{Q}^2]^5}\bigg]\\& \cdot \bigg[1+\frac{i(\mathbf{P}^2+\bar{Q}^2)}{2z(1-z)q^+}\bigg(\frac{\Delta z^+_2+\Delta z_3^+}{2}\bigg)\bigg]-\,d^{abc}\frac{i\delta^{ij}\mathbf{P}^k}{4z(1-z)q^+[\mathbf{P}^2+\bar{Q}^2]^3}|\Delta z_2^+ - \Delta z_3^+|  \Bigg\}.     
\end{split}
\end{equation*}
This cross section can be further simplified to a combination of TMD distributions as follows:
\begin{equation*}
    \begin{split}
  \frac{d\sigma_{\gamma^*_L \rightarrow q_1\bar{q}_2}}{dP.S.}&\Bigg|_{3*F^{\perp-}} \sim d^{abc}\Bigg[\frac{\delta^{ij}}{[\mathbf{P}^2+\bar{Q}^2]^4}-\frac{4\mathbf{P}^i\mathbf{P}^j\mathbf{P}^k}{[\mathbf{P}^2+\bar{Q}^2]^5}\bigg]\bigg[\Phi(0,0,0, \Delta\mathbf{b}) \\& + \color{black}\frac{\mathbf{P}^2+\bar{Q}^2}{2q^+P^-z(1-z)}\color{black}\int_{-1}^1dx\delta'(x)\Phi(x, -\frac{x}{2}, -\frac{x}{2}, \Delta \mathbf{b})\bigg] 
   \\&+ f^{abc}\frac{2i(1-2z)\mathbf{P}^i\mathbf{P}^j\mathbf{P}^k}{z(1-z)2q^+P^-[\mathbf{P}^2+\bar{Q}^2]^4} \color{black}\int_{-1}^1dx\delta'(x)\Phi(0, x, -x, \Delta \mathbf{b}) \\& +d^{abc}\frac{i\delta^{ij}\mathbf{P}^k}{4\pi z(1-z)2q^+P^-[\mathbf{P}^2+\bar{Q}^2]^3}\int_{-1}^1dx\bigg[\frac{1}{(x+i\epsilon)^2} + \frac{1}{(x-i\epsilon)^2}\bigg] \Phi(x, -\frac{x}{2}, -\frac{x}{2}, \Delta \mathbf{b}).
\end{split}
\end{equation*}
Similar results have also been obtained in the transverse photon case.

\section{Conclusions}
We obtained the DIS dijet cross section
 within the CGC formalism
 at Neik and NLP accuracy in the correlation limit, including contributions with two- and three-point gluon field correlators with transverse separation. All of these correlators can be further connected to proper definitions of two- and three-point gluon TMD distributions, respectively. It is possible as long as we take into account potential issues with operator ordering, which may become more challenging when one would like to go beyond Neik or NLP approximations, where expected will be the appearance of functions with four field insertions and so on. However, the present calculation already shows us the connection between CGC and TMD formalisms  in the case of dijet production at back-to-back limit beyond eikonal and leading power accuracy. \newline

\section*{Acknowledgments}

TA are supported in part by the National Science Centre (Poland) under the research Grant No. 2023/50/E/ST2/00133 (SONATA BIS 13). GB is supported in part by the National Science Centre (Poland) under the research Grant No. 2020/38/E/ST2/00122 (SONATA BIS 10). AC is supported in part by the National Science Centre (Poland) under the research Grant No. 2021/43/D/ST2/01154 (SONATA 17).

\end{document}